\documentstyle[prl,aps,twocolumn]{revtex}

\begin{document}

\twocolumn[\hsize\textwidth\columnwidth\hsize\csname@twocolumnfalse\endcsname


\title{Experimental improvement of chaotic synchronization due to \\
multiplicative time--correlated Gaussian noise}

\author{V.\ P\'erez-Mu\~nuzuri\footnote{E-mail: vicente@fmmeteo.usc.es; http://fmmeteo.usc.es} and M.N. Lorenzo}
\address{Group of Nonlinear Physics, Faculty of Physics, \\
University of Santiago de Compostela, 15706 Santiago de Compostela, Spain}

\date{Received: June 26, 1998; Revised: October 12, 1998}

\maketitle

\begin{abstract}
The effect of time-correlated zero--mean Gaussian noise on chaotic synchronization
is analyzed experimentally in small--size arrays of Chua's circuits.
Depending on the correlation time, an improvement of the synchronization is found
for different values of the noise amplitude and coupling diffusion between circuits.
\end{abstract}



\vspace{.15in}

]

Recently there has been considerable interest in stochastic resonance, 
i.e., the enhanced
response of a system to an external signal induced by noise 
[Wiesenfeld \& Moss, 1995; Gammaitoni, et al., 1998; Luchinsky, et al., 1998], a
phenomenon in which noise has a {\it creative\/} role. Moreover, the influence of noise has also
been studied within the context of arrays of cells 
[Lindner, et al., 1996; Braiman, et al., 1995a;1995b; Gailey, et al., 
1997; Shuai, et al., 1998]. Here, 
the coupling strength and the noise intensity play an important role for 
{\it array enhanced stochastic resonance} 
[Lindner, et al., 1995]. Applications of noise to biological
systems or in engineering problems could be of special relevance.
On the other hand, the behavior of uncoupled chaotic systems under the influence of 
external noise has been the subject of recent work 
[Maritan \& Banavar, 1994; Pikovsky, et al., 1994; 
Herzel \& Freund, 1995; Malescio, 1996; Gade \& Basu, 1996; Longa, et al., 1996; 
Shinbrot, et al., 1993; S\'anchez, et al., 1997;1999]. The main
idea behind these papers is that uncoupled chaotic systems cannot be synchronized by means
of an identical noise signal (Gaussian noise of zero mean), except for a noise with some
non--zero bias.

In this Letter, the role of a time correlated Gaussian noise
on diffusively chaotic coupled cells is 
analyzed. The dynamical noise used in this
Letter is a Gaussian noise of zero mean of the Ornstein-Uhlenbeck type 
[Sancho, et al., 1982], characterized by a correlation function
\begin{equation}
\langle \xi(t)\,\xi(t') \rangle={\alpha \over \tau} \exp \left( -\vert t-t' \vert \over \tau \right)\ 
\label{correlation}
\end{equation}
where $\tau$ is the correlation time and $\alpha$ is the noise amplitude. 
In the limit $\tau \rightarrow 0$ the white--noise limit is recovered. 
We emphasize here the case of {\em global noise}, where the noise is identical
at each site, as opposed to the case of incoherent or {\em local noise}, where the noise
is uncorrelated from site to site.

Experiments have been performed with $N$ resistively coupled 
Chua's circuits ($N=3$) 
[Madan, 1993; Chua, 1995] in the chaotic regime,
accordingly to the design introduced by
[S\'anchez, et al., 1997;1999]. 
Each circuit, $j=1,2,3$, in the array is defined by the following evolution equations,
\begin{eqnarray}
C_1 {{d\,V_{1,j}}\over{d t}}&=&{1\over R}\,(V_{2,j}-V_{1,j})-h(V_{1,j}) + \nonumber\\
& &{1 \over R_c} \left( V_{1,j+1} + V_{1,j-1} - 2\,V_{1,j} \right) \nonumber\\
C_2 {{d\,V_{2,j}}\over{d t}}&=&{1\over R} (V_{1,j}-V_{2,j})+i_{L,j} \label{chuaor}\\
L {{d\,i_{L,j}}\over{d t}}&=&-V_{2,j}-r_0\,i_{L,j} \nonumber
\end{eqnarray}
where $V_{1}$, $V_{2}$, and $i_L$, the voltages across $C_1$ and $C_2$ and
the current through $L$, respectively, are the three variables that describe
the dynamical system, resulting from a straightforward application of
Kirchhoff's law. The parameters have the following meaning:
$C_1$ and $C_2$ are the two capacitances, $L$ the inductance, $R$ the
resistance that couples the two capacitors and $r_0$ the inner
resistance of the inductor.
Circuits were connected through capacitor $C_1$ by resistances $R_c$,
leading to a diffusion term in the potential differences [Chua, 1995], with a coupling
coefficient $D \propto 1/R_c$.
Circuits at the boundaries are only connected with one neighbor.

The three-segment piecewise-linear characteristic of the nonlinear resistor
(Chua's diode) is defined by,
\begin{equation}
h(V_{1})=G_b\,V_{1}+{1\over 2} (G_a-G_b)\left[\vert V_{1}+B_p\vert-\vert V_{1}-B_p
\vert \right] 
\label{chnonlin}
\end{equation}
where $G_a$ and $G_b$ are the slopes of the inner and outer regions
of $h(V_1)$, respectively, and $B_p=1\,V$ defines the location of the
breaking points of the three-slope nonlinear characteristic $h(V_{1})$.

An experimental setup of three identical Chua's circuits driven by noise has been built. Their components
are defined by $(C_1, C_2, L, r_0, R)=(10\,{\rm nF}, 100\,{\rm nF}, 10\,{\rm mH},
20\,\Omega, 1.1\,{\rm k}\Omega)$. The slopes of the nonlinear characteristic $h(V)$,
Eq.~(\ref{chnonlin}), are defined by $G_a=-8/7000$ and $G_b=-5/7000$.
Figure~\ref{fig_setup} shows the schematic diagram of the experimental setup. 
The circuits were sampled with a digital
oscilloscope (Hewlett-Packard 54825A) with a maximum sampling rate of $4 \times 10^9$
samples per second, 1.5 GHz bandwith, and a record length of
32000 points, connected to a PC for data processing.

The external noise has been introduced multiplicatively
using a recently introduced circuit [S\'anchez, et al., 1997;1999] that enables to
drive the nonlinear element by using the voltage from an external
source. The nonlinear element is driven, in general, by the
voltage coming from an external source, not necessarily the voltage
coming from capacitor $C_1$, as may happen in the case of a
standard Chua's circuit. Thus, it is a voltage controlled
current source (VCCS) with a characteristic defined by Eq.~(\ref{chnonlin}). 
This yields the following evolution equation for the voltage across the capacitor $C_1$,
\begin{equation}
C_1\,\dot{V_1}={{V_2-V_1}\over R}-h(V_1+\xi(t))\ 
\label{multi}
\end{equation}
where it is easy to see that the noise term yields a multiplicative contribution. 

The time correlated noise, $\xi(t)$, (Eq.~(\ref{correlation})), is obtained electronically 
by passing the output voltage of a white Gaussian noise generator, $\xi_w(t)$, through a single--pole
active filter with a time constant $\tau=R_b\,C_b$, before being
applied to the circuit as it is shown in Fig.~\ref{fig_setup} 
[McClintock \& Moss, 1989; Luchinsky, et al., 1998].
The external white noise has been generated by
using a function generator (Hewlett-Packard 33120A). Their characteristics,
gaussian distribution and zero mean in the absence of an offset, have been
adequately checked\footnote{As the bandwith ($\approx 10$ MHz) of the white noise from the HP generator
is much higher than the characteristic frequencies of the Chua's circuit, 
then for our purposes, we can consider this ideal "white" noise.}. 

Figures~\ref{fig_prop}(a-d) show the physical properties of an experimental time--correlated 
noise before being added to the voltage $V_1$ in 
the Chua's circuit. The power spectrum
of the noise (b) cannot be considered to be flat within the frequency range of interest, 
$\tau^{-1} \approx 200$ Hz. For small noise
intensities, the obtained probability distribution does not fit perfectly to the Gaussian distribution as 
a consequence of the experimental noise in the sampling process (Fig.~\ref{fig_prop}(c)). The correlation function
was also calculated experimentally (d) and compared with the theoretical function given by Eq.~(\ref{correlation}).
Our realization of noise decays exponentially within a time scale almost equal to the expected 
theoretical value. 

In order to characterize the {\it degree} of synchronization between cells of the array, we
introduce the following time--averaged quantity,
\begin{equation}
K = \lim_{T \rightarrow \infty} {1 \over T} \sum_{t=1}^{T} \left( \frac{1}{N-1} 
\sum_{j=2}^{N} \left[ V_{1,j-1}^t - V_{1,j}^t \right]^{2}\right)
\label{entropy}
\end{equation}
This function is positive defined and
it is equal to zero when all the cells in the array are globally synchronized. 
Since $K$ can serve as a measure 
of the array complexity, in this context it can be related to the Kolmogorov--Sinai entropy 
[Benettin, et al., 1976; Klimontovich, 1996]. 

The main effect of a colored Gaussian noise on an array of diffusively coupled chaotic
cells is to improve the synchronization between units.
Figure~\ref{fig_main} shows the evolution of $K$ as a function of the correlation time $\tau$ for
different values of the coupling resistance $R_c$. The mean value of $K$ increases with $R_c$ 
as expected, so a scaling factor was introduced
in order to compare the different observed behaviors of $K$.
In general, independently of the specific value of $R_c$
as $\tau$ increases, first $K$ decreases almost exponentially, reaches a minimum, 
and then rises smoothly until a constant saturation value is attained for 
$\tau \gg 1$. The minimum of $K$ ($K_{min}$) corresponds to an optimum choice of $\tau$ ($\tau_{min}$) to obtain 
the best synchronization. 

For intermediate values of $\tau$, the time correlated Gaussian noise periodically modulates $V_1$ when
driving the nonlinear element. A resonance effect between the Chua's time scale and
the noise correlation time, $\tau$, should be expected, since the power spectrum
of the noise cannot be considered to be flat within the frequency range of interest, 
$\tau^{-1}$. This resonance effect could explain 
the improvement of chaotic synchronization observed in Fig.~\ref{fig_main} for $K=K_{min}$ 
[Lorenzo \& P\'erez-Mu\~nuzuri , 1999].
Here, the double--scroll attractor becomes
periodically asymmetric with increasing noise amplitude, 
as well as blurred (Figs.~\ref{fig_atractor}(a-b)), due to the slow dynamics of the noise. 

The value of $\tau$ corresponding to $K_{min}$ was found to increase with the 
coupling resistance as it is shown in Fig.~\ref{fig_taumin}. 
As $\tau \rightarrow \tau_{min}$, a stronger interaction between the
two characteristic time scales of both cell and noise should be expected, then improving the 
synchronization between circuits. For $\tau=\tau_{min}$, in terms of frequency locking,
the dynamics of the cell could be simplified to that of an oscillator forced periodically with a frequency 
equal to $\tau_{min}^{-1}$.  
Then, for a chain of linearly coupled oscillators, its dynamics can be described in terms of a plane
wave solution with a wave velocity proportional to $\sqrt{D}$. The wave dispersion relation is 
given by $\omega \propto \sqrt{D}/\lambda$, with $\omega$ the wave frequency and $\lambda$ the wave length. 
For small size arrays, it can be considered that $\lambda$ 
is fixed by the boundary conditions. In this case, the wave frequency increases
with the coupling strength. Thus, in order to obtain locking between the internal oscillation frequency 
and the external periodic forcing,
as the coupling resistance increases, the external forcing period (in our case this is related to the 
value of $\tau_{min}$) should also increase. Obviously, the explanation above is a simplification of
the problem, since the chaotic dynamics cannot be mapped in a simple way to that of an oscillator. Nevertheless,
our aim is to stress the similarity between the classical frequency locking problem that occurs in 
a chain of oscillators forced periodically and the behavior of $K$ for $\tau \approx \tau_{min}$. Here, the locking
does not occur for a single value of the frequency, but for a range of frequencies that gives rise to a wide behavior
of $K$ as a function of $\tau$ near the onset of resonance. 

On the other hand, the two limits, $\tau \rightarrow 0$ and $\tau \rightarrow \infty$, in Fig.~\ref{fig_main} deserve
further comments.
When $\tau \rightarrow 0$, the white Gaussian noise limit is recovered and circuits in the array 
do not become synchronized to each other independently of the variance of the noise 
[S\'anchez, et al., 1997;1999]. In this case, the structure of the 
unperturbed double--scroll attractor gets smeared out with the increasing noise amplitude,
corresponding to a decreasing signal-to-noise ratio, while no evidence of
synchronization behavior is observed.
Similarly, when $\tau \rightarrow \infty$ the term $\xi(t)$ in Eq.~(\ref{multi}) behaves as a constant
value added to the voltage $V_1$. Noise affects the double--scroll dynamics that becomes 
asymmetric, while no synchronization is observed between cells within the array. In fact, for high enough 
noise amplitude, the main effect will be a biased signal that will induce a regularization in the system.
This effect is analogous to that of some chaos suppression methods that achieve this result through
perturbations in the system variables [Mat\'{\i}as \& G\"u\'emez, 1994;1996].

Figure~\ref{fig_ampl} shows the dependence of $K$ with the noise amplitude for a small value of $\tau$ near
the white limit case. As expected, increasing the noise strength leads to a worse synchronization between circuits
($K$ increases). 

The influence of the number of circuits in the array, $N$, on chaotic synchronization by
time correlated noise is also being studied 
[Lorenzo \& P\'erez-Mu\~nuzuri, 1999]. 
Preliminar results show that the larger the array, the larger the needed fluctuations to
improve chaotic synchronization; i.e. the variance of the noise must
increase as $N$ increases. 

We have observed a non-monotonic dependence of the degree of synchronization measured in terms of $K$ as
a function of the noise correlation time $\tau$ and the coupling between cells in a one-dimensional array. 
In fact, for values of $\tau$ of the order of the time 
scale of the chaotic attrator
a stochastic resonance effect is found that could explain the observed minimum value of $K$. 
In other words, the effect of color noise to improve the chaotic synchronization between cells 
is more robust than that of white noise for a constant noise amplitude.

\vspace{0.1in}

We wish to thank C.\ Rico for his help with the experimental part of this
work and Profs. M.G.\ Velarde and M.A. Mat\'{\i}as for fruitful discussions. 
This work was supported by DGES and Xunta de Galicia under Research Grants 
PB97--0540 and XUGA--20602B97, respectively.


\section*{References}
\begin{description}
\item Benettin, G., Galgani, L. \& Strelcyn, J.M. [1976] "Kolmogorov entropy and numerical
experiments", {\it Phys. Rev. A} {\bf 14}(6), 2338-2345.

\item Braiman, Y., Ditto, W.L., Wiesenfeld, K. \& Spano, M.L. [1995a] "Disorder-enhaced 
synchronization",  {\it Phys. Lett. A} {\bf 206}, 54-60.

\item Braiman, Y., Lindner, J.F., \& Ditto, W.L. [1995b] "Taming spatiotemporal chaos with 
disorder", {\it Nature} {\bf 378}, 465-467.

\item Chua, L.O. [1995] "Special issue on nonlinear waves, patterns and 
spatiotemporal chaos in dynamic arrays", {\it IEEE Trans. Circuits Syst.} {\bf 42}(10).

\item Gade, P.M., \& Basu, C. [1996] "The origin of non--chaotic behavior in identically driven
systems", {\it Phys. Lett. A} {\bf 217}, 21-27.

\item Gailey, P.C., Neiman, A., Collins, J.J. \& Moss, F. [1997] "Stochastic resonance in 
ensembles of nondynamical elements: The role of internal noise",  {\it Phys. Rev. Lett.} {\bf 79}, 4701-4704.

\item Gammaitoni, L., H\"anggi, P., Jung, P. \& Marchesoni, F. [1998] "Stochastic resonance",
{\it Rev. Mod. Phys.} {\bf 70}(1), 223-287.

\item Herzel, H. \& Freund, J. [1995] "Chaos, noise, and synchronization reconsidered",
{\it Phys. Rev. E} {\bf 52}(3), 3238-3241.

\item Klimontovich, Yu. L., [1996] "Relative ordering criteria in open systems",
{\it Uspekhi Fiz. Nauk.} {\bf 166}(11), 1231-1243.

\item Lindner, J.F., Meadows, B.K., Ditto, W.L., Inchiosa, M.E. \& Bulsara, A.R. 
[1995] "Array enhanced stochastic resonance and spatiotemporal synchronization", 
{\it Phys. Rev. Lett.} {\bf 75}(1), 3-6.

\item Lindner, J.F., Meadows, B.K., Ditto, W.L., Inchiosa, M.E. \& Bulsara, A.R. 
[1996] "Scaling laws for spatiotemporal synchronization and array enhanced 
stochastic resonance", {\it Phys. Rev. E} {\bf 53}(3), 2081-2086.

\item Longa, L., Curado, E.M.F. \& Oliveira, F.A. [1996] "Roundoff--induced coalescence of
chaotic trajectories", {\it Phys. Rev. E} {\bf 54}(3), 2201-2204.

\item Luchinsky, D.G., McClintock, P.V.E. \& Dykman, M.I. [1998] "Analogue studies of nonlinear systems", 
{\it Rep. Prog. Phys.} {\bf 61}, 889-997.

\item Madan, R.N. [1993] {\it Chua's Circuit: A Paradigm for Chaos}, (World Scientific, Singapore).

\item Malescio, G. [1996] "Noise and synchronization in chaotic systems", {\it Phys. Rev. E}
 {\bf 53}, 6551-6554.
 
\item Maritan, A. \& Banavar, J.R. [1994] "Chaos, noise and synchronization", 
{\it Phys. Rev. Lett.} {\bf 72}, 1451-1454.

\item McClintock, P.V.E. \& Moss, F. [1989] "Anologue techniques for the study of 
problems in stochastic nonlinear dynamics" in {\it Noise in Nonlinear Dynamical Systems. Vol.3} edited by
F. Moss \& P.V.E. McClintock, (Cambridge Univ. Press, UK), pp 243-271.

\item Mat\'{\i}as, M.A. \& G\"u\'emez, J. [1994] "Stabilization of chaos by proportional pulses 
in the system variables", {\it Phys. Rev. Lett.} {\bf 72}, 1455-1458.

\item Mat\'{\i}as, M.A. \& G\"u\'emez, J. [1996] 
"Chaos suppression in flows using proportional pulses in the system variables", 
{\it Phys. Rev. E} {\bf 54}, 198-209.

\item Lorenzo, M.N. \& P\'erez-Mu\~nuzuri, V. [1999] "Array enhanced chaotic synchronization by colored
Gaussian noise", {\it Phys. Rev. E} {\bf 60}, 2779-2787.

\item Pikovsky, A.S. [1994] "Comment on chaos, noise and synchronization",  {\it Phys. Rev. Lett.} 
{\bf 73}(21), 2931.

\item S\'anchez, E., Mat\'{\i}as, M.A. \& P\'erez-Mu\~nuzuri, V. [1997] "Analysis of 
synchronization of chaotic systems by noise: An experimental study", {\it Phys. Rev. E} {\bf 56}, 4068-4071.

\item S\'anchez, E., Mat\'{\i}as, M.A. \& P\'erez-Mu\~nuzuri, V. [1998] "An experimental setup for
studying the effect of noise on Chua's circuit", {\it IEEE Trans.\ on Circ.\ and Syst.\ I}, {\bf 46}, 517-520.
 
\item Sancho, J.M., San Miguel, M., Katz, S.L. \& Gunton, J.D. [1982] "Analytical and
numerical studies of multiplicative noise", {\it Phys. Rev. A} {\bf 26}, 1589-1609.

\item Shinbrot, T., Grebogi, C., Ott, E. \& Yorke, J.A. [1993] "Using small perturbations to control chaos",
{\it Nature} {\bf 363}, 411-417.

\item Shuai, J.W. \& Wong, K.W. [1998] "Noise and synchronization in chaotic neural networks", {\it Phys. 
Rev. E} {\bf 57}, 7002-7007.

\item Wiesenfeld, K. \& Moss, F. [1995] "Stochastic resonance and the benefits of noise: from ice
ages to crayfish and SQUIDs", {\it Nature} {\bf 373}, 33-36.

\end{description}


\begin{figure}
\caption[]{Schematic diagram of the experimental setup used to introduce 
noise in a multiplicative way in an array of
diffusively coupled Chua's circuits. Noise is added to voltage $V_1$ and is used to drive the 
nonlinear element of the circuit
(see Eq.~(\ref{multi})). The noise is buffered from all VCCS's ensuring thereby no interaction between the circuits
except that due to the coupling resistances $R_c$. The white Gaussian noise generated by the function generator is
transformed by the high-cut filter in a Gaussian noise of zero mean of the Ornstein-Uhlenbeck type
(see Eq.~(\ref{correlation})) with a correlation time $\tau=R_b\,C_b$, 
which is then added to the signal from capacitor $C_1$ (also buffered). The low-frequency cutoff of the filter,
determined by $R_a\,C_a$, is fixed at $\approx$ 1 Hz, and the high-frequency cutoff is adjustable by tuning $R_b$ 
($C_b=1\,{\rm \mu F}$). The output noise finally passes through a variable-gain operational amplifier (not shown) 
before being applied to the circuits.}
\label{fig_setup}
\end{figure}

\begin{figure}
\caption[]{Characterization of a time-correlated Gaussian noise obtained after a white zero--mean Gaussian noise
is passed through a single--pole filter. (a) Temporal evolution of the color noise with
amplitude 500 mV (peak-to-peak) and correlation time $\tau = 5$ ms. (b) Power spectrum of the 
noise signal. (c) Probability function distribution. The line shows a fitting of the experimental 
results to a Gaussian curve. (d) Correlation function (continuous line).
The dashed line corresponds to the theoretical curve given by Eq.~(\ref{correlation}).}
\label{fig_prop}
\end{figure}

\begin{figure}
\caption[]{Log--linear plot of $K$ as a function of the time correlation $\tau$ for three
different values of the coupling resistance $R_c$. 
Since the length of the time series is determined by the record length of the oscilloscope and the time
scale of the Chua's circuits, in order to improve the statistic, 50 realizations of the noise were
carried out for each value of $\tau$. 
In doing so, we substitute the limiting value of $T \rightarrow \infty$, in Eq.~(\ref{entropy}), for $T \gg \tau$, in 
such a way that the most probable value of $K$ becomes identical to the ensemble average of $K$ 
when the number of experiments (realizations of noise) is large.
The experimental data are shown as symbols, while lines represent 
an interpolation to the previous ones: crosses ($\times$) and a dashed line
for $R_c=6.8\,\Omega$, rhombi ($\Diamond$) and a dot--dashed line for $R_c=14\,\Omega$, and
circles ($\circ$) and a solid line for $R_c=27\,\Omega$.
The obtained values of $K$ were scaled between 0 and 1 for a better representation of the phenomenon. 
The minimum and maximum values of $K$ for each coupling resistance are: 0.11 and 0.13 for $R_c=6.8\,\Omega$,
0.34 and 0.37 for for $R_c=14\,\Omega$, and 1.39 and 1.51 for $R_c=27\,\Omega$. 
Noise amplitude (peak-to-peak): 250 mV.}
\label{fig_main}
\end{figure}

\begin{figure}
\caption[]{Effect of a time--correlated noise on the 
double--scroll (chaotic) attractor (a) of the Chua's circuit for intermediate values of $\tau$. 
As $\tau$ increases, the attractor becomes 
smeared out as well it becomes "periodically" asymmetric, finally loosing the double-scroll appearence (b). 
The image presented in panel (b) was acquired after the oscilloscope was stopped at a given instant of time.
Otherwise, other possible asymmetric shapes of the attractor could have been obtained. 
Noise amplitude (peak-to-peak): (a) 10 mV and (b) 400 mV. Correlation time; $\tau = 50$ ms. $R_c = 6.8\,\Omega$.}
\label{fig_atractor}
\end{figure}

\begin{figure}
\caption[]{Dependence of the time correlation value, corresponding to the minimum of $K$, 
with the coupling resistance, $R_c$. Noise amplitude (peak-to-peak): 250 mV.}
\label{fig_taumin}
\end{figure}

\begin{figure}
\caption[]{Dependence of $K$ with the noise amplitude (peak-to-peak) of a time correlated noise. Note the linear
dependence observed for the range of used noise amplitudes. $R_c = 10\,\Omega$, $\tau=0.01$ s.}
\label{fig_ampl}
\end{figure}


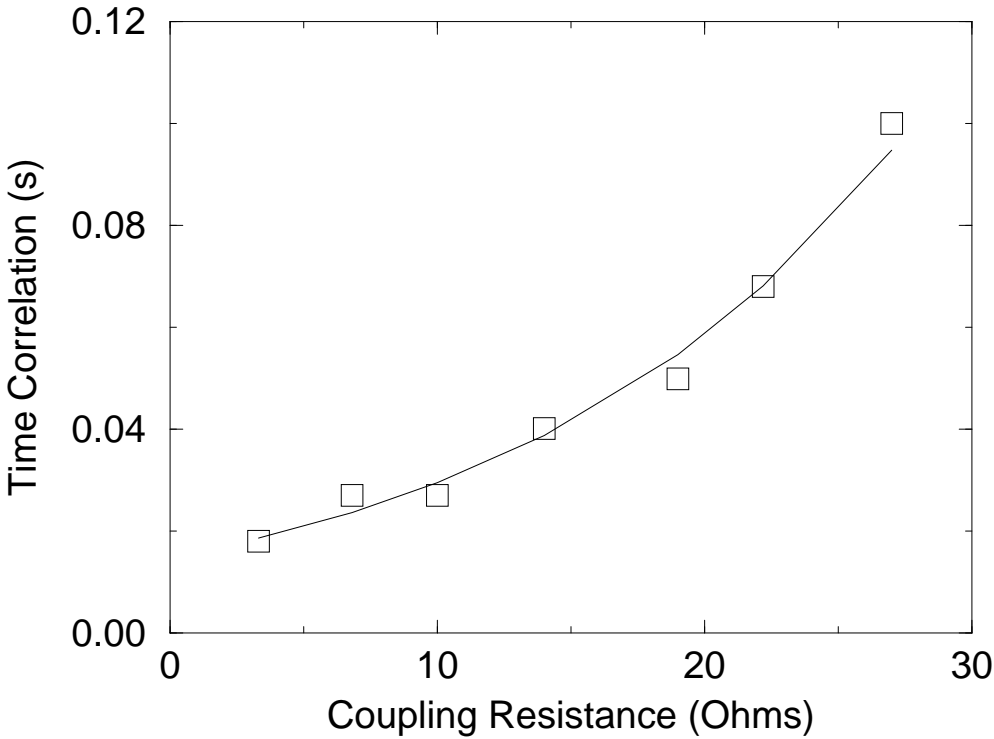
\end{document}